\documentclass[preprint,showpacs,superscriptaddress,amsmath,amssymb,showkeys]{revtex4}

\usepackage{graphicx}
\usepackage{dcolumn}
\usepackage{bm}

\begin{document}

\title{Sterile neutrino analysis of reactor-neutrino oscillation.}

\author{S. K. Kang} \email{skkang@seoultech.ac.kr}
\affiliation{School of Liberal Arts, Seoul National University
of Science and Technology, Seoul 139-743, Korea}
\affiliation{Pittsburgh Particle Physics, Astrophysics, and Cosmology Center,
Department of Physics and Astronomy, University of Pittsburgh, Pittsburgh, PA 15260, USA}
\author{Y. D. Kim} \email{ydkim@sejong.ac.kr}
\affiliation{Department of Physics, Sejong University, Seoul, 143-747, Korea}
\author{Y. Ko} \email{yjko4u@gmail.com}
\author{K. Siyeon} \email{siyeon@cau.ac.kr}
\affiliation{Department of Physics, Chung-Ang University, Seoul, 156-756, Korea}

\date{\today}
\begin{abstract}
Sterile neutrinos are one candidate to explain anomalies in neutrino oscillations. The mass-difference-driving oscillation between flavors can be probed only within specific combinations of baseline and flight energy. For a neutrino whose mass is completely unknown, it is necessary to scan all available ranges in spectrum and all accessible baselines. Here, we present four-neutrino analysis of the results announced by RENO and Daya Bay, which performed the definitive measurements of $\theta_{13}$ based on the disappearance of the inverse-beta-decay antineutrino at km-order baselines. Our results within 3+1 scheme include the exclusion curve of $\Delta m_{41}$ vs. $\theta_{14}$, and the adjustment of $\theta_{13}$ due to the contribution of $\theta_{14}$ to the disappearance of electron antineutrinos.
\end{abstract}

\pacs{11.30.Fs, 14.60.Pq, 14.60.St}
\keywords{neutrino oscillation, mixing angles, sterile neutrino}
\maketitle \thispagestyle{empty}


\section{Introduction}

On top of the solar and atmospheric neutrino experiments, a series of recent oscillation experiments, T2K\cite{Abe:2011sj}, MINOS\cite{Adamson:2011qu}, Double Chooz\cite{Abe:2011fz}, Daya Bay\cite{An:2012eh}, and RENO\cite{Ahn:2012nd}, have firmly established a framework of neutrino oscillations among three flavor neutrinos mixed with three mass eigenstates through a unitary matrix.
According to the 3$\nu$ global analysis\cite{Fogli:2012ua}, the $3\sigma$ ranges of the physical oscillation parameters are given as in the following: $6.99<\Delta m_{21}^2/10^{-5}eV^2<8.18, ~ 2.59<\sin^2\theta_{12}/10^{-1}<3.59, ~2.19(2.17)<\Delta m_{32}^2/10^{-3}eV^2<2.62(2.61), ~1.69(1.71)<\sin^2\theta_{13}/10^{-2}<3.13(3.15),$ and $3.31(3.35)<\sin^2\theta_{23}/10^{-1}<6.36(6.63)$ for normal(inverted) hierarchy. While all the three mixing angles are now known to be different from zero, the values of CP violating phases are completely unknown yet. Although there are a number of global analysis which presented the values of masses and mixing parameters consistent with among themselves \cite{Fogli:2012ua,Tortola:2012te,GonzalezGarcia:2012sz}, we focus on $\theta_{13}$ and its relatives obtained by RENO and Daya Bay.

In spite of the confirmation of three flavors of neutrinos, we do not concretely exclude the existence of new kinds of neutrinos. According to the LEP experimental result for  invisible $Z$ boson decay \cite{Beringer:1900zz}, if there exist new types of neutrinos with mass below 45 GeV, they should be sterile neutrinos which are singlet fields under the SU(2) weak interaction. Although sterile neutrinos do not interact with the electroweak gauge bosons, they can mix with three active neutrinos, leading to the oscillation between active and sterile neutrinos.

Inactive singlet neutrinos are familiar in utilizing the see-saw mechanism \cite{seesaw}, but they are untouchably heavy. On the other hand, the existence of light sterile neutrinos with masses about $\mathrm{O}(1)$ eV or less has not been phenomenologically ruled out. It is naturally considered that their existence may affect cosmology such as Big Bang Nuclearosynthesis, Cosmic Microwave Background, Hubble constant and galaxy power spectrum etc. Many literatures have studied the impact of sterile neutrinos on cosmology and obtained some constraints on the effective numbers of light neutrino species and on the sum of light neutrino masses which may favor the existence of sterile neutrinos \cite{cosmology}. The anomalies observed in the LSND \cite{Aguilar:2001ty}, MiniBooNE \cite{AguilarArevalo:2007it}, Gallium solar neutrino experiments \cite{gallium} and some reactor experiments
 \cite{Declais:1994su}, over the past several years,  also can partly be reconciled by the oscillations between active and sterile neutrinos, if more than one kind of sterile neutrino are heavier than three active neutrinos \cite{resl-sterile}
.

We examine whether the oscillation between a sterile neutrino and active neutrinos is plausible, especially by interpreting the results released from Daya Bay and RENO. For the sake of simplicity, we assume that one type of sterile neutrino is added into the contents of neutrinos. The analysis is restricted within a narrow range of $\Delta m_{41}^2$, since the setups of two experiments are optimized for $\Delta m_{31}^2$. The search for $\Delta m_{41}^2$ oscillation with a sterile neutrino is possible only if the order of $\Delta m_{41}^2$ is not far different from the order of $\Delta m_{31}^2$. There is no hope to probe a sterile neutrino in kilometer-baseline reactor neutrino oscillations such as RENO and Daya Bay, if the mass squared difference between the sterile and active neutrinos is not less than $1 \mbox{eV}^2$. Thus, our study does not necessarily cover the sterile neutrinos introduced to reconcile the anomalies observed in the short baseline experiments. In fact, efforts to search sterile neutrinos are being made with all types of oscillations with different baselines \cite{Abazajian:2012ys,Adamson:2010wi,Mention:2011rk,Bora:2012pi,Gao:2013tha,Giunti:2012tn,Conrad:2012qt,Gninenko:2011hb,Kopp:2013vaa}.
There is a work that tried to probe sterile neutrino parameters with recent reactor neutrino experiments \cite{Bora:2012pi,Gao:2013tha}, where a different range of $\Delta m_{41}^2$ is focused on, compared with this work.

This article is organized in the following outline: In Section II, the survival probability of electron antineutrinos is presented in four-neutrino oscillation scheme. We exhibit the dependence of the oscillating aspects on the order of $\Delta m_{41}^2$, when the reactor neutrinos with the energy range 1 to 8 MeV are detected after a travel along km-order baseline. In Section III, the rate-only analysis results announced by RENO and Daya Bay are re-analyzed in version of four-neutrino oscillation, and the relevant spectral shape analysis follows in the next section. In Section IV, the curves of four-neutrino oscillation are compared with the data obtained at the experiments in order to search for any clue for a sterile neutrino and in order to see the change in $\sin^22\theta_{13}$ in the coexistence with the sterile neutrinos. Broad ranges of $\Delta m_{41}^2$ and $\sin^22\theta_{14}$ remain not being excluded. The exclusion bounds of $\sin^22\theta_{14}$ and the best fit of $\sin^22\theta_{13}$ are summarized in conclusion.

\section{Four neutrino oscillation}

The three-neutrino transformation from mass basis to flavor basis is given in terms of three angles and a Dirac phase \cite{PMNS}:
    \begin{eqnarray}
    U_\mathrm{PMNS}=
    R_{23}(\theta_{23})R_{13}(\theta_{13},\delta_1)R_{12}(\theta_{12}),
    \end{eqnarray}
where $R_{ij}(\theta_{ij})$ denotes the rotation of the i-j block by the angle $\theta_{ij}$.
When a 3+1 model is assumed as the minimal extension, the unitary transformation from the mass basis of $\{m_1,~m_2,~m_3,~m_4\}$ to the flavor basis $\{\nu_e,~\nu_\mu,~\nu_\tau,\nu_s\}$ is given in terms of six angles and three Dirac phases:
    \begin{eqnarray}
    \widetilde{U}_\mathrm{F} &=& R_{34}(\theta_{34})R_{24}(\theta_{24},\delta_2)R_{14}(\theta_{14}) \cdot \nonumber \\
    & & \cdot R_{23}(\theta_{23})R_{13}(\theta_{13},\delta_1)R_{12}(\theta_{12},\delta_3).
    \label{4by4trans}
    \end{eqnarray}
The 4-by-4 $\widetilde{U}_\mathrm{F}$ is expressed as
\begin{widetext}
    \begin{eqnarray}
    \widetilde{U}_\mathrm{F} &=&
	\left(\begin{matrix}	
        c_{14} & 0 & 0 & s_{14} \\
		-s_{14}s_{24} & c_{24} & 0 & c_{14}s_{24} \\
		-c_{24}s_{14}s_{34} & -s_{24}s_{34} & c_{34} & c_{14}c_{24}s_{34} \\
		-c_{24}c_{34}s_{14} & -s_{24}c_{34} & -s_{34} & c_{14}c_{24}c_{34}
	    \end{matrix}\right)
    \left( \begin{matrix}
    U_{e1} & U_{e2} & U_{e3} & 0 \\
    U_{\mu1} & U_{\mu2} & U_{\mu3} & 0 \\
    U_{\tau1} & U_{\tau2} & U_{\tau3} & 0 \\
    0 & 0 & 0 & 1
        \end{matrix} \right)
         \\
    &=& \left( \begin{matrix}
    c_{14}U_{e1} & c_{14}U_{e2} & c_{14}U_{e3} & s_{14} \\
    \cdots & \cdots & \cdots & c_{14}s_{24} \\
    \cdots & \cdots & \cdots & c_{14}c_{24}s_{34} \\
    \cdots & \cdots & \cdots & c_{14}c_{24}c_{34}
        \end{matrix} \right),
\end{eqnarray}
\end{widetext}
where the PMNS type of 3-by-3 matrix $U_\mathrm{PMNS}$ with three rows, $(U_{e1} ~ U_{e2} ~ U_{e3}), ~ (U_{\mu1} ~ U_{\mu2} ~ U_{\mu3})$ and $(U_{\tau1} ~ U_{\tau2} ~ U_{\tau3}),$ is imbedded. The CP phases $\delta_2$ and $\delta_3$ introduced in Eq.(\ref{4by4trans}) are omitted for simplicity, since they do not affect the electron antineutrino survival probability at the reactor neutrino oscillation.

The survival probability of $\bar{\nu}_e$ produced from inverse beta decay is
    \begin{eqnarray}
    P_\mathrm{th}(\bar{\nu}_e\rightarrow\bar{\nu}_e) &=& |\sum_{j=1}^4|\widetilde{U}_{ei}|^2 \exp{i\frac{\Delta m_{j1}^2L}{2E_\nu}}|^2 \\
        &=& 1- \sum_{i<j}4|\widetilde{U}_{ei}|^2|\widetilde{U}_{ej}|^2\sin^2(\frac{\Delta m_{ij}^2L}{4E_\nu}),
    \end{eqnarray}
where $\Delta m_{ij}^2$ denotes the mass-squared difference $m_i^2-m_j^2$.
In the limit where $|\widetilde{U}_{e3}|$ and $|\widetilde{U}_{e4}|$ are much smaller than one, the survival probability of $\bar{\nu}_e$ is determined by the following three terms (as long as $m_4$ is much larger than the others):
    \begin{eqnarray}
    P_\mathrm{th}(\bar{\nu}_e\rightarrow\bar{\nu}_e) &=& 1-c_{14}^4c_{13}^4\sin^2 2\theta_{12}\sin^2(1.27\Delta m_{21}^2\frac{L}{E}) \nonumber \\
    &-& c_{14}^4\sin^2 2\theta_{13}\sin^2(1.27\Delta m_{31}^2\frac{L}{E})  \label{pee} \\
    &-& \sin^2 2\theta_{14}\sin^2(1.27\Delta m_{41}^2\frac{L}{E}). \nonumber
    \end{eqnarray}
The oscillation pattern of $P_\mathrm{th}$ as $L/E$ increases in a logarithmic way is described in Fig.\ref{fig1:distance}, where three patterns of oscillating probabilities are shown according to the order of $\Delta m_{41}^2$. It is shown that the order of $\Delta m_{41}^2$ to probe a 4th neutrino must be not much larger than that of $\Delta m_{31}^2$, since both RENO and Daya Bay have baselines optimized for $\Delta m_{31}^2$. The first bump in each curve corresponds to the oscillation due to $\Delta m_{41}^2$, while the second bump that appears near $4\sim5 \mathrm{m/MeV}$ corresponds to the oscillation due to $\Delta m_{31}^2$. If $\Delta m_{41}^2$ is much less than $\mathcal{O}(0.01)\mathrm{eV}^2$, the amplitude of the $\Delta m_{41}^2$ oscillation can appear only in the superposition with the $\Delta m_{31}^2$ oscillation. The bound on $\Delta m_{41}^2$ at which the two types of bumps can separate is about $0.008\mathrm{eV}^2$. On the other hand, if $\Delta m_{41}^2 > \mathcal{O}(0.03)$, the oscillating aspect of the far-to-near ratio becomes unclear. Hereafter, we consider the mass of the sterile neutrino $m_4$ within $0.008\mathrm{eV}^2 < \Delta m_{41}^2 < 0.05\mathrm{eV}^2$.

\section{RENO and Daya Bay: Reactor neutrino oscillation experiments }

\begin{figure}
\resizebox{80mm}{!}{\includegraphics[width=0.75\textwidth]{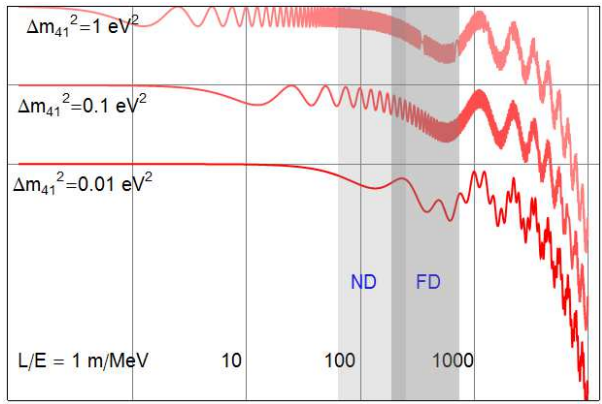}}
\caption{\label{fig1:distance}The 4-neutrino oscillating aspects of $P_\mathrm{th}$ in Eq.(\ref{pee}) according to the baseline-to-energy ratio, $L/E$, for different orders of $\Delta m_{41}^2$. The shaded areas ND and FD denote the distances to the near detector and the far detector divided by the energy range 2 to 8 MeV.}
\end{figure}

The six baselines of the near detector(ND) and the far detector(FD) of RENO are $L_\mathrm{near}(\mathrm{meters})=\{ 660, 445, 302, 340, 520, 746\}$ and $L_\mathrm{far}(\mathrm{meters})=\{1560, 1460, 1400, 1380, 1410, 1480\}$, while their flux-weighted averages $\overline{L}_\mathrm{near}$ and $\overline{L}_\mathrm{far}$ are 407.3m and 1443m, respectively. The baselines of Daya Bay, named EH1, EH2, and EH3, have lengths of EH1=494m, EH2=554m, and EH3=1628m, so that, conventionally, EH1 and EH2 are regarded as near detectors while EH3 is regarded as a far detector. Since the energy of the reactor neutrinos falls mostly in the range of 2 MeV to 8 MeV, the typical $L/E$ for reactor neutrino oscillation experiments like RENO and Daya Bay is estimated to be between 60 km/MeV and 270 km/MeV for the ND, and between 180 km/MeV and 800 km/MeV for the FD. The coverage of the detectors of RENO and Daya Bay is described in Fig.\ref{fig1:distance}.
\begin{figure*}
\resizebox{80mm}{!}{\includegraphics[width=0.75\textwidth]{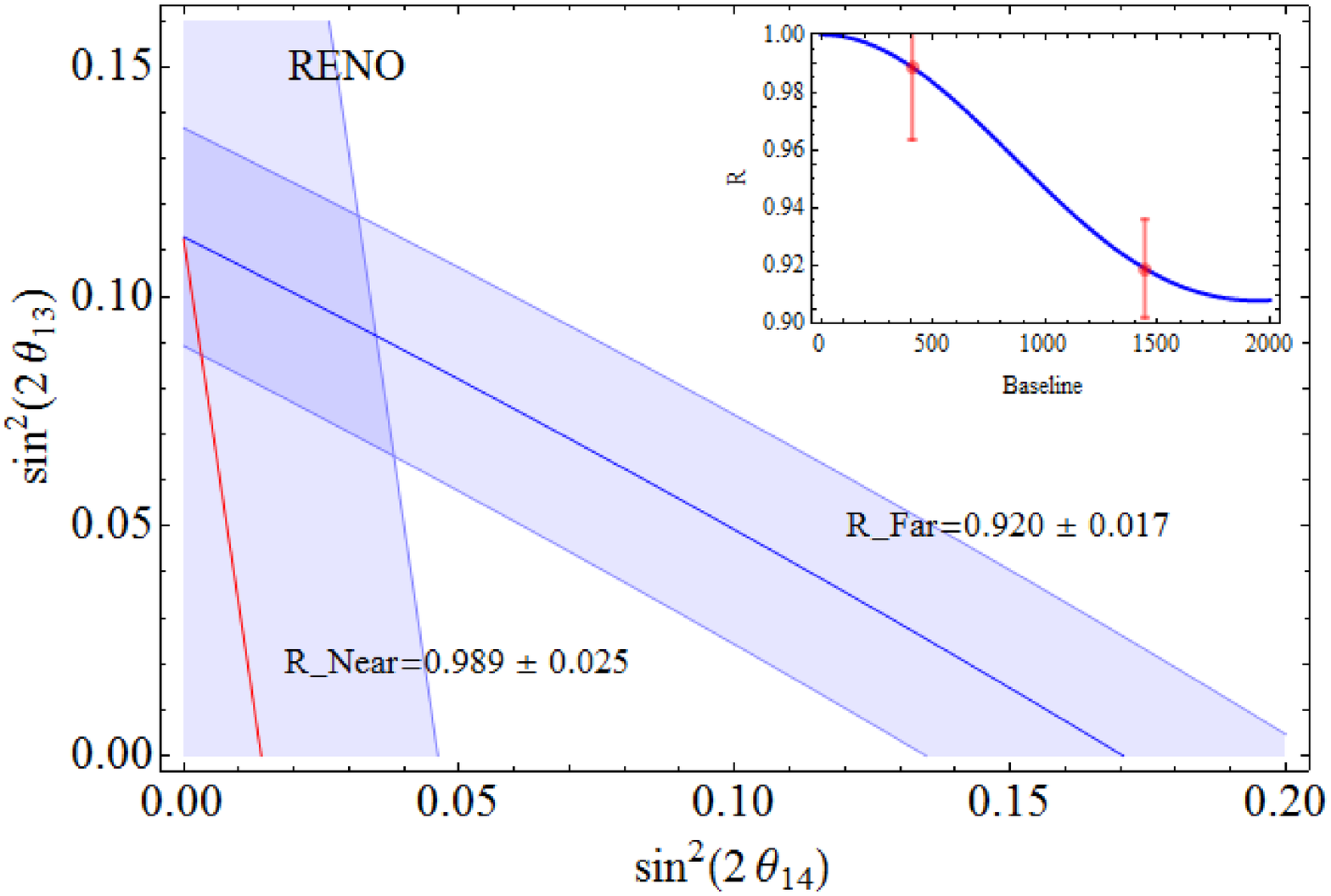}}
\resizebox{80mm}{!}{\includegraphics[width=0.75\textwidth]{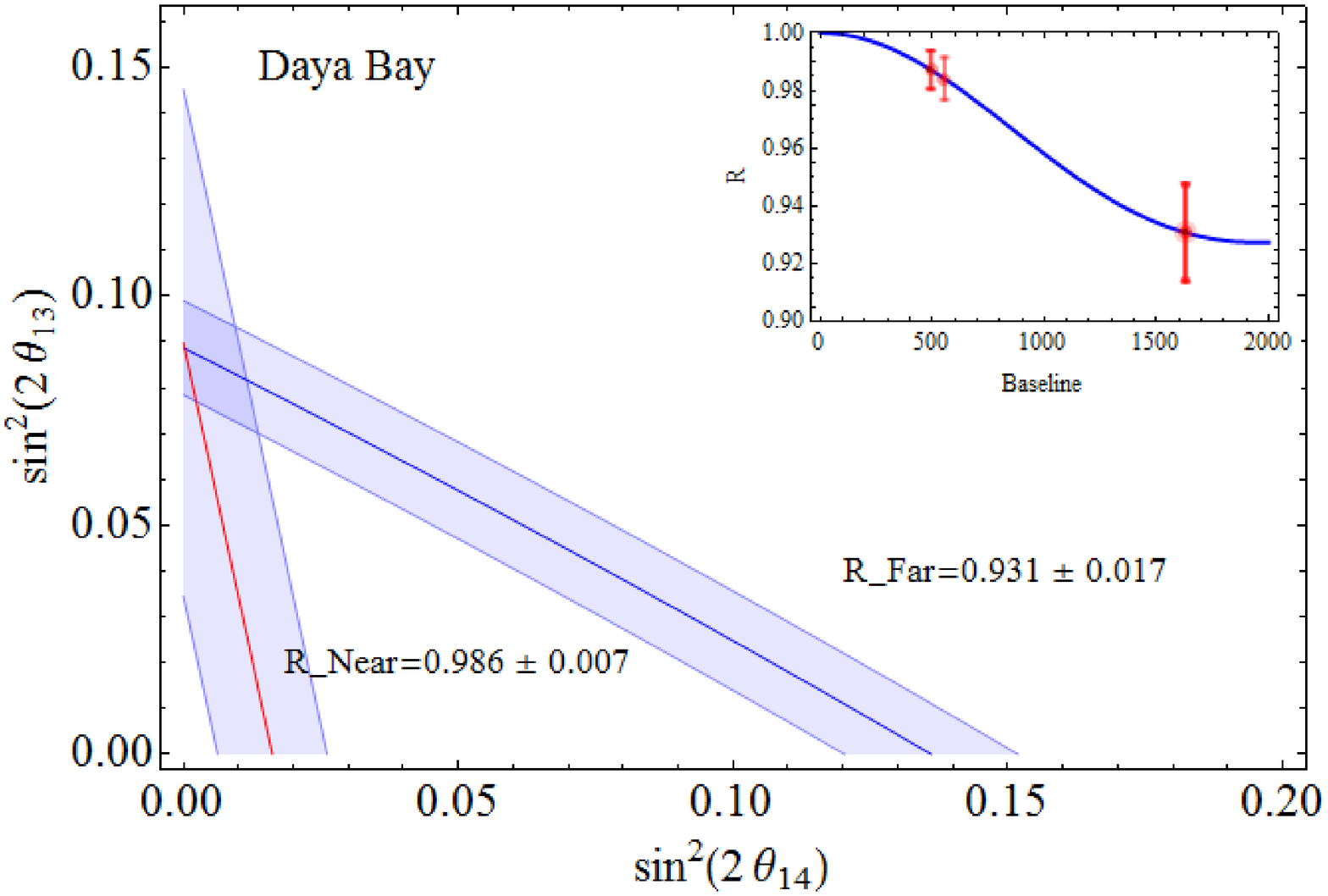}}
\caption{\label{fig:theta13theta14}
Four-neutrino analysis for the observed to expected ratios at both ND and FD: The ranges of $R_\mathrm{near}$ and $R_\mathrm{far}$ in the insets released by RENO and Daya Bay are reinterpreted in terms of the availability in $\sin^22\theta_{13}$ and $\sin^22\theta_{14}$ in a four-neutrino oscillation. }
\end{figure*}

The rate-only analysis of neutrino oscillation takes the average over accessible energies of the neutrinos emerging from the reactors. The measured probability of survival is
    \begin{eqnarray}
    \langle P \rangle = \frac{\int P_{\mathrm{s}}(E) \sigma_{\mathrm{tot}}(E) \phi(E) dE}{\int  \sigma_{\mathrm{tot}}(E) \phi(E) dE}, \label{avg_prob}
    \end{eqnarray}
where $\sigma_{\mathrm{tot}}(E)$ is the total cross section of inverse beta decay(IBD), and $\phi(E)$ is the neutrino flux distribution from the reactor. For Reno, the survival probability $P_{\mathrm{s}}$ in Eq.(\ref{avg_prob}) at each detector is evaluated as $P_{\mathrm{near}}=0.0678 P_{n1} + 0.1493 P_{n2} + 0.3419 P_{n3} + 0.2701 P_{n4} + 0.1150 P_{n5} + 0.0558 P_{n6}$ and $P_{\mathrm{far}}=0.1373 P_{f1} + 0.1574 P_{f2} + 0.1809 P_{f3} + 0.1856 P_{f4} + 0.1780 P_{f5} + 0.1608 P_{f6}$, based on the relative distances. Each $P_{ni}$ or $P_{fi}$ is given as  $P_\mathrm{th}$ in Eq.(\ref{pee}). The total cross section of IBD is given as \cite{Vogel:1999zy,Mueller:2011nm}
    \begin{eqnarray}
    \sigma_{\mathrm{tot}}(E)=0.0952 \left( \frac{E_e\sqrt{E_e^2-m_e^2}}{1\mathrm{MeV}^2}\right)\times 10^{-42}\mathrm{cm}^2,
    \end{eqnarray}
where $E_e=E_\nu-(M_n-M_p)$. The flux distribution $\phi(E)$ from the 4 isotopes $(\mathrm{U}^{235},\mathrm{Pu}^{239},\mathrm{U}^{238},\mathrm{Pu}^{241})$ at the reactors  is expressed by the following exponential of a 5th order polynomials of $E_\nu$
    \begin{eqnarray}
    \phi(E_\nu)=\exp\left(\sum_{i=0}^5 f_i E_\nu^i \right), \label{flux}
    \end{eqnarray}
where $f_0=+4.57491 \times 10, f_1=-1.73774 \times 10^{-1}, f_2=-9.10302 \times 10^{-2}, f_3=-1.67220 \times 10^{-5}, f_4=+1.72704 \times 10^{-5}$, and $f_5=-1.01048 \times 10^{-7}$ are obtained by fitting the total flux of the four isotopes with the fission ratio expected at the middle of the burn up period of the reactors \cite{Huber:2011wv}. Including the product of $\sigma_\mathrm{tot}(E_\nu)$ and $\phi(E_\nu)$ in the integrand in Eq. (\ref{avg_prob}) results in average probability curve $\langle P \rangle$ shown in the inset of Fig. \ref{fig:theta13theta14}. When $\sin^2 2\theta_{13}=0.113$ (as announced by Reno), $\theta_{14}=0$ curve is consistent with the released result.

On the other hand, the probability at Daya Bay is evaluated by three groups of six reactors, EH1(AD1 \& AD2), EH2(AD3), and EH3(AD1, AD2, \& AD3), which catch neutrinos from three groups of reactors, D1 \& D2, L1 \& L2, and L3 \& L4. The survival probability at each detector from different reactors is evaluated as follows: $P(EH1)=0.795P_{A1}+0.143P_{B1}+0.062P_{C1}$, $P(EH2)=0.065P_{A2}+0.512P_{B2}+0.423P_{C2}$, and $P(EH3)=0.246P_{A3}+0.379P_{B3}+0.375P_{C3}$, where $P_{Ai}, P_{Bi}$ and $P_{Ci}$ are the probabilities at the i-th group of detectors due to reactors, D1 \& D2, L1 \& L2, and L3 \& L4, respectively. The coefficients are determined according to the relative distances. Substituting $(P(EH1)+P(EH2))/2$ into $P_S$ in Eq.(\ref{avg_prob}) gives the survival probability at near detectors. Likewise, substituting $P(EH3)$ into $P_S$ gives the probability at the far detectors. Although the flux distribution at Daya Bay should have been obtained independently, $\phi(E_\nu)$ for RENO in Eq.(\ref{flux}) was included in the integration and, fortunately, the $\theta_{14}=0$ plot appears consistent with the published curve for $\sin^2 2\theta_{13}=0.089$ from Daya Bay, as shown in the inset of Fig.\ref{fig:theta13theta14}.

In rate-only analysis, the value of $\sin^2 2\theta_{13}$ is determined by obtaining the curve of $\langle P \rangle$ in Eq.(\ref{avg_prob}) which can match the measured ratios $R$s of the observed to expected flux at each detector. However, there is a technical difference in the definition of the expected flux using RENO's approach compared to Daya Bay's. The definition of $R$ that each experiment used is discussed in details in the next section.

The main figures in Fig.\ref{fig:theta13theta14} interpret the measurements of the far-to-near ratio in terms of a four-neutrino oscillation. Two shaded bands in each frame indicate the ratios of the measured to expected at ND and FD, where the expected flux is an estimation of the flux expected without neutrino oscillation. The two bands in the plane of $\sin^2 2\theta_{13}$ and $\sin^2 2\theta_{14}$ for the RENO experiment have ratios of $R_\mathrm{far}=0.920 \pm 0.017$ at FD and $R_\mathrm{near}=0.989 \pm 0.025$ at ND. The intersection indicates the combination of $\sin^2 2\theta_{13}$ and $\sin^2 2\theta_{14}$ allowed by the error bars in the inset. Meanwhile, the two bands for the Daya Bay experiment showed ratios of                                                                                             $R_\mathrm{far}=0.931 \pm 0.017$ at FD and $R_\mathrm{near}=0.986 \pm 0.007$ at ND; likewise, the intersection indicates the allowed combination of the two angles by the error bars in the inset. Depending on the existence of the 4th neutrino and the magnitude of $\sin^22\theta_{14}$, the value of $\sin^22\theta_{13}$ might have a different value from that previously determined, as shown in Fig. \ref{fig:theta13theta14}. Such a combination of $\sin^2 2\theta_{13}$ and $\sin^2 2\theta_{14}$ will be examined more carefully within spectral shape analysis.

\section{Shape Analysis}

\begin{figure*}
\resizebox{180mm}{!}{\includegraphics[width=0.75\textwidth]{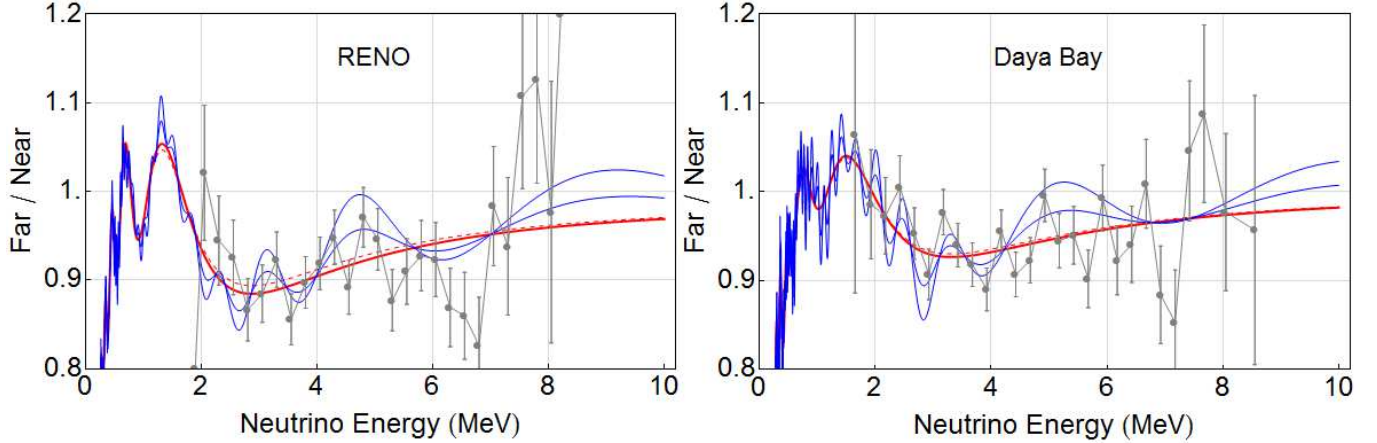}}
\caption{\label{fig:shape_analysis}
The curves of the ratio of the survival at FD to the survival at ND of IBD antineutrinos: Red represents the $\Delta m_{31}^2$ oscillation in the three-neutrino analysis, while the blue represent the superposition of two types of oscillations: one for $\Delta m_{31}^2=0.00232 \mathrm{eV}^2$ and the other for $\Delta m_{41}^2=0.016 \mathrm{eV}^2$. The data and errors in each figure are reproductions of original data. For the RENO(Daya Bay) setup, the red curve is drawn by Eq. (\ref{far2near})(Eq. (\ref{far2near_db})) with $\sin^2 2\theta_{13}=0.124 (0.0936)$ and the red dashed uses $\sin^2 2\theta_{13}=0.113 (0.089)$. The blue curves representing the superposition of two oscillations are obtained by $\sin^2 2\theta_{14}=0.05$ and 0.1, respectively.}
\end{figure*}

\subsection{RENO}

One of the RENO's first results was the ratio of the observed to the expected number of antineutrinos in the far detector, $R=0.920 \pm 0.017$ (see Ref.\cite{Ahn:2012nd}), where the observed is simply the number of events at FD. On the other hand, the expected number of events at FD can be obtained through several adjustments of the number of events at ND, i.e.,
    \begin{eqnarray}
    R &\equiv&\frac{[\mathrm{Observed ~at ~FD}]}{[\mathrm{Expected ~at ~FD}]} \label{obs2exp}\\
     &\equiv& \frac{[\mathrm{No. ~of ~events ~at ~FD}]}{[\mathrm{No. ~of ~events ~at ~ND}]^*} ~, \label{far_to_near}
    \end{eqnarray}
where the number of events at each detector is normalized. The normalization of the neutrino fluxes at ND and FD requires an adjustment between the two individual detectors which includes corrections due to DAQ live time, detection efficiency, background rate, and the distance to each detector. The numbers of events at FD and ND in Eq.(\ref{far_to_near}) have already been normalized by these correction factors, and so we have $R_\mathrm{far}=0.920\pm0.017$ and $R_\mathrm{near}=0.989\pm0.025$ as shown in Fig.\ref{fig:theta13theta14}. The normalization guarantees $R=1$ at the center of reactors. RENO gets rid of the oscillation effect at ND when evaluating the expected number of events at FD by dividing the denominator of Eq. (\ref{far_to_near}) by 0.989 which is taken from $R_\mathrm{near}$. Now,
    \begin{eqnarray}
    R = \frac{[\mathrm{No. ~of ~events ~at ~FD}]}{[\mathrm{No. ~of ~events ~at ~ND}] ~/~0.989} ~, \label{jungsik}
    \end{eqnarray}
In rate-only analysis, the ratio of the observed to the expected number of events at FD in Eq.(\ref{obs2exp}) is just the survival at FD, since the denominator in Eq. (\ref{jungsik}) is eliminated. Thus, $R$ coincides with the $R_\mathrm{far}$ in Fig. \ref{fig:theta13theta14}.

In spectral shape analysis, however, the denominator cannot be neglected, since the oscillation effect at ND differs depending on the neutrino energy. The data points in Fig.\ref{fig:shape_analysis} are obtained by the definition of the ratio $R$ given in Eq. (\ref{obs2exp}) and Eq. (\ref{far_to_near}) per 0.25MeV bin, as the energy varies from 1.8MeV to 12.8MeV. The theoretical curves overlaid over the data are also obtained parallel to the ratio in Eq.(\ref{jungsik}). They are described by
\begin{eqnarray}
\frac{P_\mathrm{th}(L_\mathrm{far})}{P_\mathrm{th}(L_\mathrm{near})(0.989)^{-1}} ~, \label{far2near}
\end{eqnarray}
where $P_\mathrm{th}(L)$ is given in Eq.(\ref{pee}). While the thick red curve is
a typical $\Delta m_{31}^2$-dominant oscillation at $\theta_{14}=0$, the blue curves are examples of a superposition of a $\Delta m_{31}^2$ oscillation and a $\Delta m_{41}^2$ oscillation. The red curve describes the oscillation related to Eq. (\ref{far2near}) when $\sin^22\theta_{13}=0.113$, as determined from the rate-only analysis. However, if the data are compared with the curve of the ratio in Eq. (\ref{far2near}) at $\theta_{14}=0$, the minimum of $\chi^2$ is obtained at $\sin^2 2\theta_{13}=0.124$, which is slightly different from $\sin^2 2\theta_{13}=0.113$, the result of the rate-only analysis. The $\Delta\chi^2$ with respect to $\sin^2 2\theta_{13}$ is drawn in Fig. \ref{fig:th13th14}(c).

\begin{figure*}
\resizebox{160mm}{!}{\includegraphics[width=0.75\textwidth]{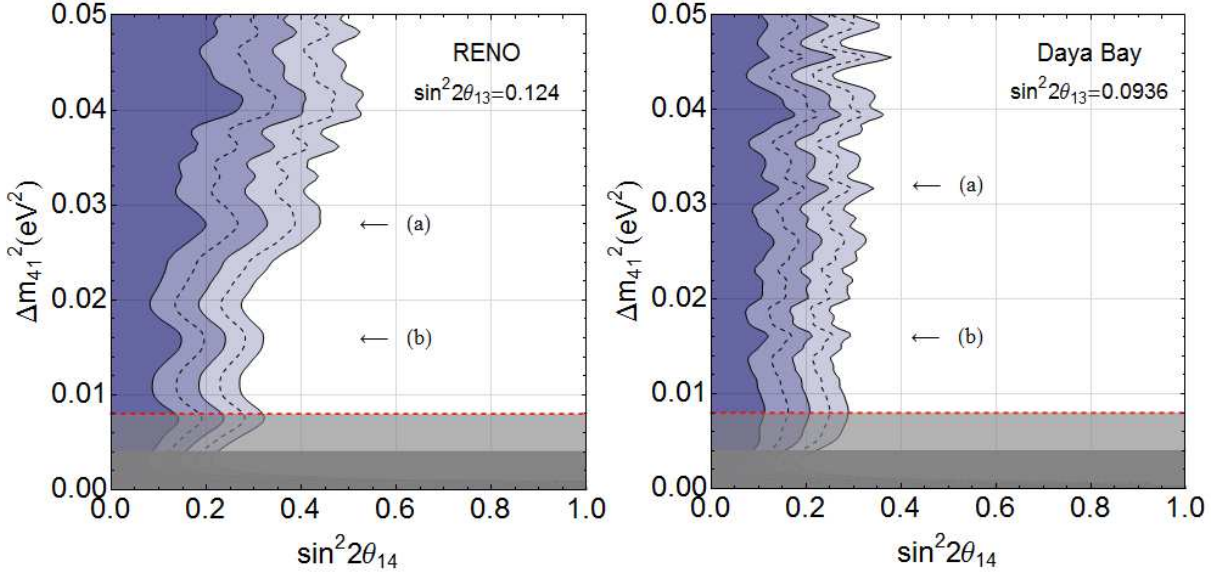}}
\caption{\label{fig:exclusion}
The $1\sigma, ~2\sigma$ and $3\sigma$ exclusion curves. Due to the limits of the baselines and the neutrino energy, $\Delta m^2_{41}$ larger than 0.05$\mathrm{eV}^2$ is excluded from the analysis. The range below 0.008$\mathrm{eV}^2$ is blocked for $\Delta m^2_{41}$, since oscillations in that range cannot be detected with a km-order baseline.  For both, broad ranges of $\sin^2 2\theta_{14}$ apparently remain unexcluded.}
\end{figure*}

\begin{figure*}
\resizebox{160mm}{!}{\includegraphics[width=0.75\textwidth]{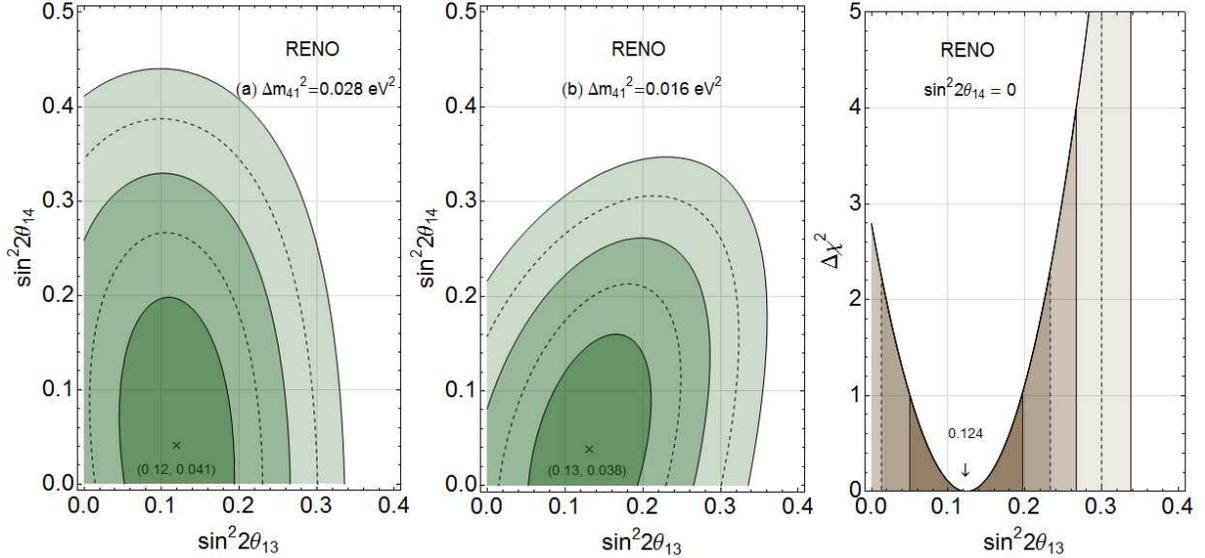}}
\caption{\label{fig:th13th14}
Four-neutrino analysis in the $\sin^2 2\theta_{13}$ and $\sin^2 2\theta_{14}$ plane for chosen values of $\Delta m^2_{41}$: (a) 0.028$\mathrm{eV}^2$ and (b) 0.016$\mathrm{eV}^2$. The last figure shows the best fit of $\sin^2 2\theta_{13}$, 0.124, and the standard deviation without $\sin^2 2\theta_{14}$.}
\end{figure*}

Even when nonzero $\theta_{14}$ is considered, the accessible range of $\Delta m_{41}^2$ to catch the 4th neutrinos at RENO or Daya Bay is very narrow, ( just above $\Delta m_{31}^2=0.00232 \mathrm{eV}^2$), since the baselines are optimized for a $\Delta m_{31}^2$ oscillation. The interpretation of the data points in Fig. \ref{fig:shape_analysis} in terms of the combined oscillations of $\Delta m_{31}^2$ and $\Delta m_{41}^2$ can be expressed as in Fig. \ref{fig:exclusion}. As expected, the $\Delta m_{41}^2$ above $0.03\mathrm{eV}^2$ is difficult to analyze in the four-neutrino oscillation scheme. The three curves in the figure correspond to the three exclusion curves at  $1\sigma, ~2\sigma$ and $3\sigma$ CLs as the result of shape analysis, i.e., with 99.7\% confidence, values of $\sin^2 2\theta_{14}$ larger than 0.3 are excluded. Figure \ref{fig:th13th14} describes the preference of the combination of $\sin^2 2\theta_{13}$ and $\sin^2 2\theta_{14}$ for a given value of $\Delta m_{41}^2$. Each figure contains $1\sigma, ~2\sigma$ and $3\sigma$ standard deviations, and the values of $\Delta m_{41}^2$ for figures (a) and (b) were chosen from a couple of crests. Figure (c) shows a slope of $\Delta \chi^2$ with respect to $\sin^2 2\theta_{13}$ at $\sin^2 2\theta_{14}=0$, which can be compared with the results of the three-neutrino rate-only analysis.

\subsection{Daya Bay}

In many aspects, the first results released by Daya Bay and RENO are parallel. There is a slight difference in what is expressed by the ratio $R$ and the data points in Fig. \ref{fig:shape_analysis} between the two experiments. Daya Bay also includes the result that $R=0.944 \pm 0.008$ (the ratio of the observed to the expected number of antineutrinos assuming no oscillations at the far detector), which implies that
    \begin{eqnarray}
    R \equiv \frac{[\mathrm{No. ~of ~events ~at ~FD}]}{[\mathrm{No. ~of ~events ~at ~ND}]} ~, \label{far2near_ratio}
    \end{eqnarray}
where the numbers of events at both detectors have been normalized for corrections including DAQ live time, detection efficiency, background rate, and distance to each detector. In comparison with RENO's $R$ from Eq.(\ref{far_to_near}), the oscillation factor at ND is not eliminated from the number of events in order to give rise to the expected number at FD. In other words, the expected number at FD takes the number of events at ND directly, while the observed at FD is the number of events at FD. Relative to the flux emerging from the reactors, the events at ND and FD are $R_\mathrm{near}=0.986\pm0.007$ and $R_\mathrm{far}=0.931\pm0.017$, so that
    \begin{eqnarray}
    R=\frac{R_\mathrm{far}}{R_\mathrm{near}}=0.944~,
    \end{eqnarray}
which was the result released from rate-only analysis.

\begin{figure*}
\resizebox{160mm}{!}{\includegraphics[width=0.75\textwidth]{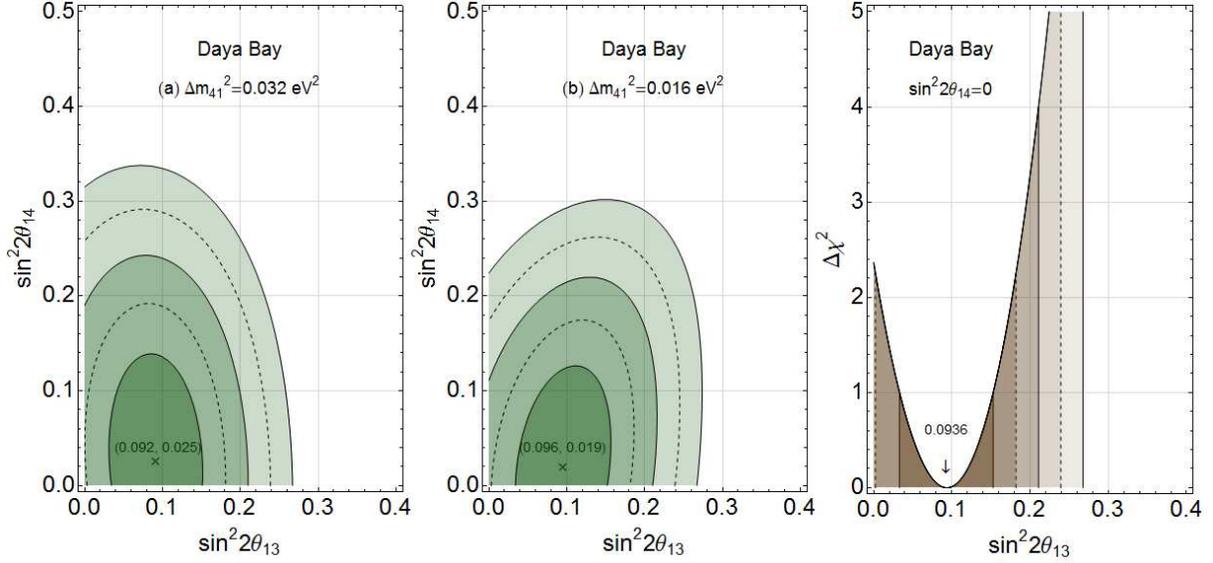}}
\caption{\label{fig:th13th14_daya}
Four-neutrino analysis in the $\sin^2 2\theta_{13}$ and $\sin^2 2\theta_{14}$ plane for chosen values of $\Delta m^2_{41}$: (a) 0.032$\mathrm{eV}^2$ and (b) 0.016$\mathrm{eV}^2$. The last figure shows the best fit of $\sin^2 2\theta_{13}$, 0.0936, and its standard deviation without $\sin^2 2\theta_{14}$. }
\end{figure*}

The data points in Fig. \ref{fig:shape_analysis} also gives $R$ in Eq.(\ref{far2near_ratio}) per 0.25MeV bin, as the energy varies from 1.8MeV to 8.5MeV.
The theoretical curves overlaid for comparison with the data are consistent with the ratio in Eq.(\ref{far2near_ratio}), and can be described by
    \begin{eqnarray}
    \frac{P_\mathrm{th}(L_\mathrm{far})}{P_\mathrm{th}(L_\mathrm{near})} ~, \label{far2near_db}
    \end{eqnarray}
which can be compared with the plots of Eq. (\ref{far2near}) for RENO.
While the thick red curve is
a typical $\Delta m_{31}^2$-dominant oscillation without the 4th neutrino, the blue curves are examples of the superposition of a $\Delta m_{31}^2$ oscillation and a $\Delta m_{41}^2$ oscillation. The red curve describes Eq.(\ref{far2near_db}) when $\sin^22\theta_{13}=0.089$ as determined from the rate-only analysis in Ref.\cite{An:2012eh}, and is actually a reproduction of the curve from Ref.\cite{An:2012eh}. However, if the data are compared with the curve of Eq.(\ref{far2near_db}) at $\theta_{14}=0$, the minimum of $\chi^2$ is obtained at $\sin^2 2\theta_{13}=0.094$, which is slightly different from $\sin^2 2\theta_{13}=0.089$, the result of rate-only analysis. The $\Delta\chi^2$ with respect to $\sin^2 2\theta_{13}$ is drawn in Fig.\ref{fig:th13th14_daya}(c).

When nonzero $\theta_{14}$ is considered, the interpretation of the data points in Fig. \ref{fig:shape_analysis} in terms of the combined oscillation of $\Delta m_{31}^2$ and $\Delta m_{41}^2$ can be expressed as in Fig.\ref{fig:exclusion}. As in RENO, $\Delta m_{41}^2$ above $0.02\mathrm{eV}^2$ is avoided in the four-neutrino oscillation schemes. The three exclusion curves corresponding to  $1\sigma, ~2\sigma$ and $3\sigma$ CLs are given from the result of spectral shape analysis, i.e., with 99.7\% confidence, values of $\sin^2 2\theta_{14}$ larger than 0.3 are excluded. For a given value of $\Delta m_{41}^2$, the combination of $\sin^2 2\theta_{13}$ and $\sin^2 2\theta_{14}$ is analyzed in Fig. \ref{fig:th13th14_daya}, where each figure contains $1\sigma, ~2\sigma$ and $3\sigma$ standard deviations. In figures (a) and (b), the values of $\Delta m_{41}^2$ are chosen to be 0.032$\mathrm{eV^2}$ and 0.016$\mathrm{eV^2}$ for a couple of crests in Fig.\ref{fig:exclusion}, respectively. Figure (c) shows the slope of $\Delta \chi^2$ with respect to $\sin^2 2\theta_{13}$ at $\sin^2 2\theta_{14}=0$, where $\Delta \chi^2$ is minimized at $\sin^2 2\theta_{13}=0.0936$.

\section{Conclusion}

If a fourth type of neutrino has a mass not much larger than the other three, the results of reactor neutrino oscillations like RENO, Daya Bay, and Double Chooz can be affected by the fourth state. For detectors established for oscillations driven by $\Delta m_{31}^2=0.00232\mathrm{eV}^2$, clues of the fourth neutrino can be perceived only if the order of $\Delta m_{41}^2$ is not much larger than that of $\Delta m_{31}^2$. Therefore, this work examined the possibility to find a kind of sterile neutrino for the range of mass-squared difference below $0.05\mathrm{eV}^2$. On the other hand, value of $\Delta m_{41}^2$ below $0.008\mathrm{eV}^2$ is not considered either, so that an approximation was used in evaluating oscillation probabilities. Otherwise, the contribution of $\Delta m_{41}^2$ oscillation to the disappearance of IBD neutrinos is difficult to separate from that of $\Delta m_{31}^2$ oscillation. We examined the two announced results of RENO and Daya Bay, in terms of a four-neutrino oscillation for a certain range of $\Delta m_{41}^2$.

The first results released by the two experiments are rate-only analyses providing the far-to-near ratio of properly normalized events. The ratios, $R=0.920\pm0.017$ from RENO and $R=0.944\pm0.008$ from Daya Bay were interpreted as $\sin^22\theta_{13}=0.113\pm0.023$ and $\sin^22\theta_{13}=0.089\pm0.011$, respectively, in their original three-neutrino analysis. For comparison, we interpret the ratios with respect to a four-neutrino oscillation. The range of $\sin^22\theta_{13}$ broadens along with the range of $\sin^22\theta_{14}$, as shown in Fig. \ref{fig:theta13theta14}, which took (and shifted) the errors from the $R$'s.

Although the spectral shape analysis was not presented due to lack of data in the first releases of RENO and Daya Bay, we included the shape analysis of each oscillation as shown in Fig.\ref{fig:shape_analysis}. Because of the accessibility of the baseline, $\Delta m^2_{41}$ larger than $0.05\mathrm{eV}^2$ is excluded. As expected, only $\Delta m^2_{41}$ below $0.03\mathrm{eV}^2$ exhibits modest oscillatory aspects for both in Fig.\ref{fig:exclusion}. Regarding RENO, Fig. \ref{fig:shape_analysis}, which examines the exclusion boundary in the $\Delta m^2_{41}$-$\sin^22\theta_{14}$ plane, $\Delta m^2_{41}=0.042\mathrm{eV}^2$ and $0.028\mathrm{eV}^2$ seem the most likely for arbitrary $\sin^22\theta_{14}$. When $\Delta m^2_{41}=0.042\mathrm{eV}^2(0.028\mathrm{eV}^2)$, values of $\sin^22\theta_{14}$ larger than 0.43(0.53) are excluded by $3\sigma$ CL as shown in Fig.\ref{fig:th13th14}. Regarding Daya Bay, there is no preferred $\Delta m^2_{41}$ and values of $\sin^2 2\theta_{14}>0.38$ are excluded by  $3\sigma$ CL as shown in Fig.\ref{fig:exclusion}. When $\sin^22\theta_{14}$ is taken into consideration, the $\chi^2$ minimum is given rise to by $\sin^2 2\theta_{13}=0.124$ for RENO and $\sin^2 2\theta_{13}=0.0936$ for Daya Bay, which are slightly different from the rate-only analysis results of the first announcements. In this rough estimation, $\sin^22\theta_{13}=0$ is not excluded by $3\sigma$ CL for either RENO or Daya Bay.

\begin{acknowledgments}
This work was supported by the National Research Foundation of Korea(NRF) grants funded by the Korea Government of the Ministry of Education, Science and Technology(MEST) (No. 2011-0003287, 2012-0004311).
\end{acknowledgments}

\appendix

\end{document}